\documentclass[pra,twocolumn,amsmath,amssymb,superscriptaddress,showpacs,numbers]{revtex4-1}
\usepackage{graphicx,psfrag,color,hyperref}
\usepackage{amssymb,latexsym,mathrsfs,amsmath}
\usepackage{xcolor}

\usepackage{braket}
\begin{document}

%%%%%%%%%%%%%%%%%%%%%%%%%%%%%% Textclass specific LaTeX commands.

\newtheorem{thm}{\protect\theoremname}
\newtheorem{defn}{\protect\definitionname}
\newtheorem{cor}{\protect\corollaryname}

\def\bea{\begin{eqnarray}}
\def\eea{\end{eqnarray}}

%\@ifundefined{date}{}{\date{}}

%\makeatother

  \providecommand{\definitionname}{Definition}
\providecommand{\theoremname}{Theorem}
\providecommand{\corollaryname}{Corollary}

\title{How to Suppress Dark States in Quantum Networks and Bio-Engineered Structures}

\author{T. P. Le}
\affiliation{\mbox{Dept. of Physics and Astronomy}, University College London, Gower Street, London WC1E 6BT, UK.}
\author{Ludovica Donati}
\affiliation{\mbox{Dept. of Mechanical Engineering}, via S. Marta 3, 50139 Florence, Italy.}
\author{Simone Severini}
\affiliation{\mbox{Dept. of Computer Science}, University College London, Gower Street, London WC1E 6BT, UK and \mbox{Institute of Natural Sciences}, Shanghai Jiao Tong University, No. 800 Dongchuan Road, Minhang District, Shanghai 200240, China.}
\author{Filippo Caruso}
\affiliation{\mbox{LENS, QSTAR}, and \mbox{Dept. of Physics and Astronomy}, via G. Sansone 1, I-50019 Sesto Fiorentino, Italy.}

%$^2$ \mbox{Department of Industrial Engineering, University of Florence,} via S. Marta 3, I-50139 Florence, Italy \\
%$^3$ \mbox{Department of Information Engineering, University of Florence,} via S. Marta 3, I-50139 Florence, Italy \& \mbox{CSDC, University of Florence, and INFN,} Sezione di Firenze, Italy}

%\date{}                     %% if you don't need date to appear
%\setcounter{Maxaffil}{0}
%\renewcommand\Affilfont{\itshape\small}

\begin{abstract}
Transport across quantum networks underlies many problems, from state
transfer on a spin network to energy transport in photosynthetic complexes.
However, networks can contain \emph{dark subspaces} that block the
transportation, and various methods used to enhance transfer on quantum
networks can be viewed as equivalently avoiding, modifying, or destroying
the dark subspace. Here, we exploit graph theoretical tools to identify the dark subspaces and show that \emph{asymptotically almost surely} they do not exist for large networks, while for small ones they can be suppressed by properly perturbing the coupling rates between the network nodes. More specifically, we apply these results to describe the recently experimentally observed and robust transport behaviour of the electronic excitation travelling on a genetically-engineered light-harvesting cylinder (M13 virus) structure. We believe that these mainly topological tools may allow us to better infer which network structures and dynamics are more favourable to enhance transfer of energy and information towards novel quantum technologies.
\end{abstract}

\date{\today}
\maketitle

\section{Introduction}

Understanding the mechanisms of optimal transport of various quantities, such as energy or information, across some underlying topology is fundamental to many problems in physics and beyond (see, for instance, Refs. \citep{Kay_2010,Mulken_2011,Mohseni_2013} and references therein). Networks can be used to model quantum channels:
for examples, states can be transferred along spin chains \citep{Christandl_2004,Aharonov_1993}. In these studies, the aim is perfect state transfer
and there tends to be a fixed Hamiltonian that drives the transfer.
Controllability of networks asks what kind of possibly time-dependent
interactions\textemdash which then affects the connectivity structure
of the network\textemdash will enable any state to be transferred \citep{Pemberton_Ross_2010}. More recently, quantum network theory
has also been applied to model how energy is transferred through biological photosynthetic complexes \citep{Caruso_2009,Engel_2007,Panitchayangkoon_2010,Panitchayangkoon_2011,Scholes_2012,Chin_2010,Caruso_2010_LHC,Olaya-Castro_2008} and over more abstract complex networks \citep{Caruso_2014,Novo_2015,Novo_2016}.
There are numerous factors that need to be considered in order to achieve optimal transport: the dynamics of the network and the approximations used,
the initial preparation and its coherence, the location of the target
node, site energies, static disorder, noise, dissipation, etc. In this context optimally refers to several transport features as absence of losses, short required time, and robustness (regardless of sudden changes of working conditions). One
hindrance to optimal transport is represented by the presence of \emph{dark} or \emph{invariant subspaces/states} \citep{Caruso_2009}. Inspired by the similar use of the term
``dark states'' in quantum optics \citep{Arimondo_1976} and condensed
matter physics \citep{Brandes_2000,Emary_2007}, Ref. \citep{Caruso_2009}
defines them as Hamiltonian eigenstates that have no overlap with
the ``target'' node on the network. They, hence, act as a
trap on the network blocking transport. Then, transport efficiency
can be increased by either avoiding the dark subspace, or applying
certain techniques to nudge states out of the dark subspace, or by
destroying the subspace \citep{Li_2015, Caruso_2016, Viciani_2015, Viciani_2016}.
Here, we will discuss these different methods to enhance quantum transport by means of graph theoretical tools, and apply them to describe the energy transport behaviour that has been recently experimentally observed for a bio-engineered light-harvesting complex realized on a cylinder (M13 virus) structure \citep{Park_2016}.

This paper is structured as follows. In Sec. \ref{sec:Quantum-Network},
we formally introduce the network, its dynamics and the corresponding
dark subspace. Sec. \ref{sec:Methods-To-Enhance} reviews methods
that are used to enhance (energy) transfer on quantum networks through
the lens of dark states: initialisation outside of the dark subspace,
using control fields, and coupling with the environment thus introducing
noise and disorder. In Sec. \ref{sec:Analysis-Of-Dark}, we employ
graph theoretical results in order to find two results on the dark subspaces on graphs: that there exist dynamics having no associated dark subspace, and that very large graphs asymptotically almost surely have no dark subspace. In Sec. \ref{sec:Environment-Coupling} we describe some applications of these studies to light-harvesting complexes. Finally, in Sec.\ref{sec:robustness} we illustrate the results numerically by changing the underlying topology of a particular system inspired by a recent experiment with genetically-engineered light-harvesting structures \citep{Park_2016}. We also highlight the importance of dephasing noise to enhance the transmission efficiency. Some conclusions are drawn in Sec. \ref{sec:Conclusion}.

\section{Quantum Network\label{sec:Quantum-Network}}

A quantum network consists of an underlying graph, on which the dynamics
is described via quantum mechanics, as opposed to the usual transition
matrices or hopping dynamics of classical networks \cite{Lovas_1993}. A graph, $G=\left(V,E\right)$,
consists of a set of vertices or nodes $V\left(G\right)$ and a
set of edges $E\left(G\right)$. Let $N=\left|V\left(G\right)\right|$
be the number of nodes on the graph. The graph can be described by
its adjacency matrix $A\left(G\right)$, defined as
\begin{align}
\left[A\left(G\right)\right]_{ij} & =\begin{cases}
\alpha_{ij}, & \text{if }\left(i,j\right)\in E\left(G\right);\\
0, & \text{otherwise},
\end{cases}
\end{align}
where $i,j\in V$ are nodes of the network, and $\alpha_{ij}$ are the weights of the edges. We consider the edges
to be undirected and without loops, unless specified
otherwise.
The coherent dynamics is described by the Hamiltonian:
\begin{align}
H_{0} & =\sum_{i=1}^{N}\hbar\omega_{i}\sigma_{i}^{+}\sigma_{i}^{-}+\sum_{i\neq j}\hbar\left[A\left(G\right)\right]_{ij}\left(\sigma_{i}^{+}\sigma_{j}^{-}+\sigma_{j}^{+}\sigma_{i}^{-}\right),
\end{align}
where $\sigma_{i}^{+}$ and $\sigma_{i}^{-}$ are the raising and
lowering operators at node $i$ respectively, $\hbar\omega_{i}$ is
the local site energy, and $\left[A\left(G\right)\right]_{ij}=\alpha_{ij}$ determines the hopping rate
(interaction) between joined nodes $i$ and $j$. In the following we will consider the single excitation approximation, as often use for light-harvesting complexes and for quantum states and information transfer \citep{Caruso_2009, Bose_2009, Caruso_2010}. Hence, the state $\ket{i}$ denotes the presence of one excitation in node $i$, \emph{i.e.} $\sigma_{i}^{+}=\ket{i}\bra{0}$,
\emph{etc.}
The exit or target node can be thought of as the location from which
a decay process transfers irreversibly excitation to a sink, labelled as $N+1$. If the target node is node $N$,
then this decay can be formally described by the addition of the Lindblad
superoperator
\begin{eqnarray}
&&\mathcal{L}_{sk}\left(\rho\right) = \\
&&\Gamma_{N+1}\left(2\sigma_{N+1}^{+}\sigma_{N}^{-}\rho\sigma_{N}^{+}\sigma_{N+1}^{-}-\left\{ \sigma_{N}^{+}\sigma_{N-1}^{-}\sigma_{N+1}^{+}\sigma_{N}^{-},\rho\right\} \right), \nonumber
\end{eqnarray}
where $\rho$ describes the state of the network, $\Gamma_{N+1}$
is the decay rate to the sink, and $\left\{ A,B\right\} =AB+BA$ is
the anti-commutator. The transmission efficiency is given
by the probability of population transfer to the final node:
\begin{align}
p_{sink}\left(t\right) & =2\Gamma_{N+1}\int_{0}^{t}\rho_{NN}\left(s\right)ds.
\end{align}
Formally, the transfer efficiency represents the probability for the electronic excitation to be transferred to the sink, while $1-p_{sink}\left(t\right)$ corresponds to the \emph{energy trapped} in the network.

Now, we consider the following definition of dark subspace \citep{Caruso_2009}:
\begin{defn}
Consider a graph $G$ with Hamiltonian dynamics $H_{0}$
and target node $N$, corresponding to the state $\ket{N}=\left(0,0,\ldots,0,1\right)$
in the site basis. The \emph{dark subspace} is the vector space spanned
by the eigenvectors of $H_{0}$ that are orthogonal to $\ket{N}$.
\end{defn}
In order to determine the dark states, it is necessary to know
the spectrum of the Hamiltonian and the position of the exit node.
The term ``dark state'' in this context was first used by Ref.
\citep{Caruso_2009}, who called the dark subspace as the ``invariant
subspace'', since it is invariant under the dynamics described above. We can also define the corresponding
\emph{light subspace} as being spanned by all the eigenvectors of
$H_{0}$ that are not orthogonal to the target node $\ket{N}$. In this last set of eigenvectors it is possible to identify a particular subset made of vectors whose scalar product with the target node is bounded by a very small positive quantity $\varepsilon$. We can define these vectors as \emph{quasi-dark states} because they are quasi-orthogonal to $\ket{N}$, hence they are not able to trap the excitation as dark states do, but they can cause transport to be very slow. Here, we introduce a new quantity, named \emph{darkness strength}, ($\varepsilon$), which enables us to quantify the eigenvector capacity in trapping the excitation inside its eigenspace: it is zero for dark states and very close to zero for quasi-dark states.

In the case of noisy quantum dynamics, that is the network is coupled
to some environment, then there can be also dissipative and dephasing processes. They can be described by the following
Lindblad superoperators,
\begin{align}
\mathcal{L}_{diss}\left(\rho\right) & =\sum_{j=1}^{N}\Gamma_{j}\left(2\sigma_{j}^{-}\rho\sigma_{j}^{+}-\left\{ \sigma_{j}^{+}\sigma_{j}^{-},\rho\right\} \right), \label{Lsuperoperetor}\\
\mathcal{L}_{deph}\left(\rho\right) & =\sum_{j=1}^{N}\gamma_{j}\left(2\sigma_{j}^{+}\sigma_{j}^{-}\rho\sigma_{j}^{+}\sigma_{j}^{-}-\left\{ \sigma_{j}^{+}\sigma_{j}^{-},\rho\right\} \right),
\end{align}
where $\Gamma_{j}$ and $\gamma_{j}$ are dissipation and dephasing
rates for node j, respectively. The total evolution of the state of the network is then
\begin{eqnarray}
&&\dot{\rho}\left(t\right)=\\
&&-\dfrac{i}{\hbar}\left[H_0,\rho\right]+\mathcal{L}_{sk}\left(\rho\right)+\mathcal{L}_{diss}\left(\rho\right)+\mathcal{L}_{deph}\left(\rho\right)\equiv\mathcal{L}\left[\rho\right], \nonumber
\label{Lindblad}
\end{eqnarray}
where $\mathcal{L}$ is the Lindblad superoperator that describes the coherent and incoherent part of the system evolution.
%\footnote{Some subsequent figures have been realized using a different model, described in \cite{Caruso_2014}, qualitatively equivalent to the model above.}.

\subsection{Examples of dark subspaces\label{subsec:Classifications-For-Common}}

In the homogeneous case of equal local energies and uniform coupling rates, the Hamiltonian $H_0$ in the first excitation
subspace is the adjacency matrix of the underlying network.
Thus, the dark subspaces of the Hamiltonian are the eigenspaces of
the network that are orthogonal to $\left(0,\ldots,0,1\right)$. Non-degenerate
eigenvalues with eigenvectors of form $\left(\ldots,0\right)$ lead to
one-dimensional dark subspaces, while eigenvalues with degeneracy $k$
are related to dark subspaces of at least dimension $k-1$, depending on whether
or not the eigenspace is entirely orthogonal to the target node --- see Appendix in Ref\citep{Caruso_2009} to see how to find them.

We can
consider the more general question of whether a network has any potential
dark subspaces\textemdash whether it has any eigenvectors with zero
entries in the site basis. Clearly, networks with degenerate eigenvalues will automatically have
dark subspaces relative to any node of the network. In terms of substructures, it has been found that $0$ and $-1$ eigenvalues are
related to stars and cliques on the network \citep{Dorogovtsev_2003,Goh_2001,Kamp_2005},
suggesting that graphs with many stars or cliques will have degenerate
$0$ or $-1$ eigenvalues, respectively.

Now, we look at some examples of dark subspaces on paths, lattice graphs and complete graphs \citep{Caruso_2009,Caruso_2014}. However, by exploiting the knowledge of the eigenspectrum of numerous other classes
of graphs \citep{Van_Mieghem_2009},
our statements about the corresponding
dark subspaces can be generalized to other complex networks.

\begin{itemize}

\item{Path and Lattice Graphs:}
State transfer on spin chains and spin networks have been studied
in the literature (\emph{e.g.}, \citep{Christandl_2004,Kay_2010}),
and they are one of the fundamental models in physics. Underling spin
chains with nearest neighbour coupling are path graphs. The eigenvalues
of path graphs are all nondegenerate $\lambda_{k}=2\cos\left(\pi k/\left(N+1\right)\right)$
for $k=1,\ldots,N$. The corresponding unnormalised eigenvectors $x_{k}$
have components $\left(x_{k}\right)_{m}=\sin\left(\pi mk/\left(N+1\right)\right)$
and zeros emerge at ``symmetry points'' that split the path graph
into equal parts \citep{Van_Mieghem_2009}. Thus, if our target node
is at any one of these zeroes, then there is a dark subspace. However,
in typical state transfer on spin chains, the target node is the end
node, where there is never a zero: hence perfect state transfer is
clearly possible because there is no relevant dark subspace. Larger lattice graphs also have dark nodes at symmetry points of the network
\citep{Caruso_2014, Martin-Hernandez_2012}.

\item{Complete Graphs:}
A fully connected network (FCN), or complete graph, of $N$ nodes, is defined as a network where there is a link between any pair of nodes. There is one eigenstate $\ket{\phi}=\left(1/\sqrt{N}\right)\sum_{j=1}^{N}\ket{j}$
with eigenvalue $\lambda_{1}=N-1$, and a degenerate
eigenspace of dimension $N-1$ with eigenvalue $\lambda_{2}=\ldots=\lambda_{N}=-1$ \citep{Van_Mieghem_2009},
whose basis can be chosen as $\ket{\psi_{j}}=\ket{1}-\ket{j}$ for
$j=2,\ldots,N$. The dark subspace is spanned by $\left\{ \ket{\psi_{j}}:j=2,\ldots N-1\right\} $,
which has dimension $N-2$. If the initial state is localised on a single node, then it is unavoidable that a component of it will lie in the
dark subspace \citep{Caruso_2009}.

\end{itemize}

\section{How To Enhance Transfer\label{sec:Methods-To-Enhance}}

In this section, we review several tools that can be exploited in order to increase the network transfer efficiency. One could either choose specific
initial states, as in Subsec. \ref{subsec:Intelligent-Initialisation},
or use control fields to time-dependently change the effective Hamiltonian
dynamics as in Subsec. \ref{subsec:Control-Fields}. Subsec.
\ref{subsec:Disorder-And-Dephasing} considers the case where disorder and dephasing are applied to the system dynamics.

\subsection{Smart Initialisation\label{subsec:Intelligent-Initialisation}}

The evolution of the eigenstates in dark subspace is
coherent and stationary (up to a phase), hence it will never lead to a state with a non-vanishing component on site $N$, \emph{i.e.} without reaching the exit node $N$. Indeed, the evolution of the dark subspace
as a whole is also invariant. If the initial state on the network
has any non-zero component in the dark subspace, that component remains
within the dark subspace and thus forever trapped on the network.
Only the components in the corresponding light subspace will transfer
to the exit. By initialising completely \emph{outside} the dark subspace, \emph{i.e.} with an initial state that is orthogonal to the dark subspace,
full transfer of the energy can occur in the limit of time $t\rightarrow\infty$.

\begin{figure}
\begin{centering}
\includegraphics[width=0.48\textwidth]{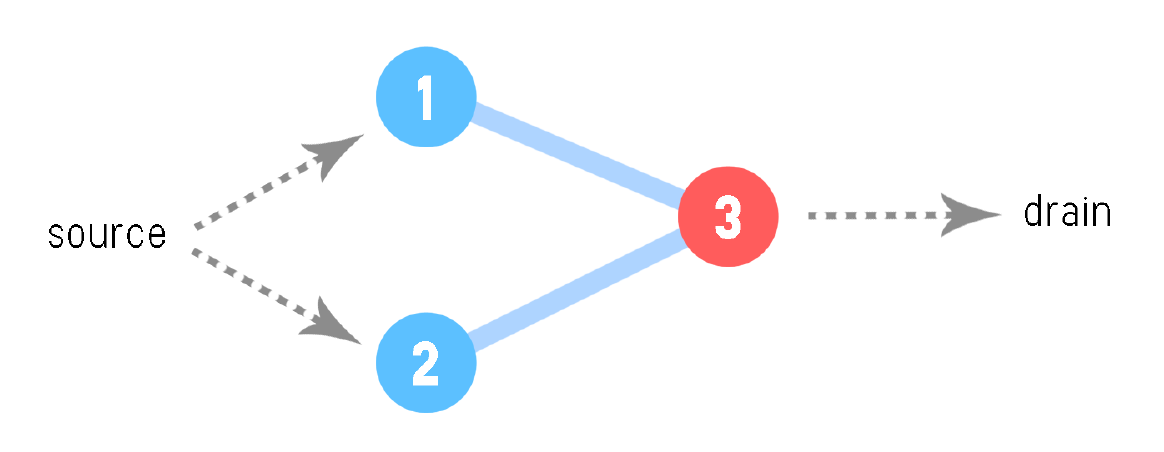}
\par\end{centering}
\caption{Trimer Configuration. The source can either initialise the state as an incoherent mixture of $\ket{1}$ and $\ket{2}$ or as  a coherent superposition of $\ket{1}$ and $\ket{2}$.}
\label{fig:Trimer-Configuration-Schijven}
\end{figure}

This line of attack is pursued by \citep{Caruso_2009, Schijven_2012}, who consider
small networks with three nodes known as trimers, shown in Fig. \ref{fig:Trimer-Configuration-Schijven}.
Trimers have one dark state that causes excitations to get trapped
\citep{Caruso_2009,Emary_2007,Michaelis_2006,Brandes_2005,M_lmer_1993,Agliari_2010}.
In fact, one can consider the following Hamiltonian (in the
first excitation subspace),
\begin{align}
H & =\begin{bmatrix}1 & 0 & 1\\
0 & 1 & 1\\
1 & 1 & 1
\end{bmatrix},
\end{align}
with the target node being $\ket{3}=\left(0,0,1\right)$. Hence, the
dark state is $\ket{D}=\left(\ket{1}-\ket{2}\right)/\sqrt{2}$, and
the other two eigenstates are $1/2\left(\ket{1}+\ket{2}\right)\pm\sqrt{2}/2\ket{3}$.
If the network state is initialised as $\ket{1}$ or $\ket{2}$,
or in an incoherent combination, then the state is inevitably partly
trapped in the dark state. Conversely, if the initial state is the
coherent superposition $\left(\ket{1}+\ket{2}\right)/\sqrt{2}$, then
perfect transfer occurs. For the in-between initialisation $\left(\ket{1}+e^{i\phi}\ket{2}\right)/\sqrt{2}$,
there is imperfect transfer, with zero transfer when $e^{i\phi}=-1$
(\emph{i.e.} initialisation as the dark state). There, dephasing in conjunction with smart initialisation
(cf. Subsec. \ref{subsec:Disorder-And-Dephasing}) is required to suppress the
dark state. This holds
for more general networks\textemdash if the initial state is completely
within the light subspace then the asymptotic transport efficiency is unity.

However, since eigenstates tend to be delocalized and a generic initial superposition will necessarily have a non-zero component in the dark subspace, other techniques will be exploited later to enhance transport.

\subsection{Control Fields\label{subsec:Control-Fields}}

Applying various control fields on the network during the transfer
process could alter the direction of the evolution of the state of
the network, and increase transport efficiency by modifying the nature
of the dark subspace.

Given a controlled system, a state $\rho^{\prime}$ is reachable from
state $\rho_{0}$ if there is a sequence of control fields (along
with any underlying Hamiltonian evolution) that will evolve $\rho_{0}$
into $\rho^{\prime}$ in some finite time. A system is \emph{controllable}
(or \emph{fully controllable} \citep{DAlessandro_2007}) if any state
in the state space is reachable from any other state \citep{Pemberton_Ross_2010,Dong_2010}.
Formally, if $H_0$ is the system/network Hamiltonian (that is time-independent),
$H_{m}$ are a set of Hamiltonians that can be applied onto the network,
and $f_{m}\left(t\right)$ are the time-varying controls, then the
total Hamiltonian under which the system evolves is
\begin{align}
H\left(t\right) & =H_0+\sum_{m}H_{m}f_{m}\left(t\right).
\end{align}
A system is fully controllable if the Lie algebra rank condition is
true: if the Lie algebra generated by $iH$ and $iH_{m}$ is isomorphic
to unitary group $u\left(N\right)$ \citep{DAlessandro_2007}, generating
all possible unitaries.

\citet{Pemberton_Ross_2010} find that the more \emph{symmetric} a
network is, the larger the dark subspace tends to be; by adding controls,
modifying the Hamiltonian etc., these symmetries can be broken and
some dark states can be accessed. In Ref. \citep{Manzano_2014},
symmetry breaking is used to make a controlled
quantum thermal switch. When the switch is ``off'', the central
qubits are all in the dark subspace and no energy can be transferred
from one side to the other. Ref. \citep{Zimbor_s_2013} breaks time-reversal
symmetry to increase transport efficiency. More generally, Refs. \citep{Polack_2009,Sander_2009}
study how symmetries of the Hamiltonian relate to lack of full controllability,
and Ref. \citep{Zeier_2011} finds that lack of certain symmetries
of the Hamiltonians are necessary for full controllability.

Control fields could also take the network into a higher excitation
subspace. By doing so, \citet{Pemberton_Ross_2010} define two
grades of dark states: weaker dark states that become non-dark by
the introduction of extra excitations or energy-preserving control
fields; and \emph{truly dark states} that require permutation symmetry-breaking\footnote{Permutation symmetry-breaking would involve unequal control fields
or disorder such that the interchange of previously indistinguishable
qubits is now invalid.} to be destroyed. As such, the weaker dark states could be used as \emph{storage},
since they are more protected from decay (from the sink) than the
non-dark states, and are more accessible than the truly dark states
\citep{Pemberton_Ross_2010}.

\begin{figure*}[!t]
\begin{centering}
\includegraphics[width=0.478\textwidth]{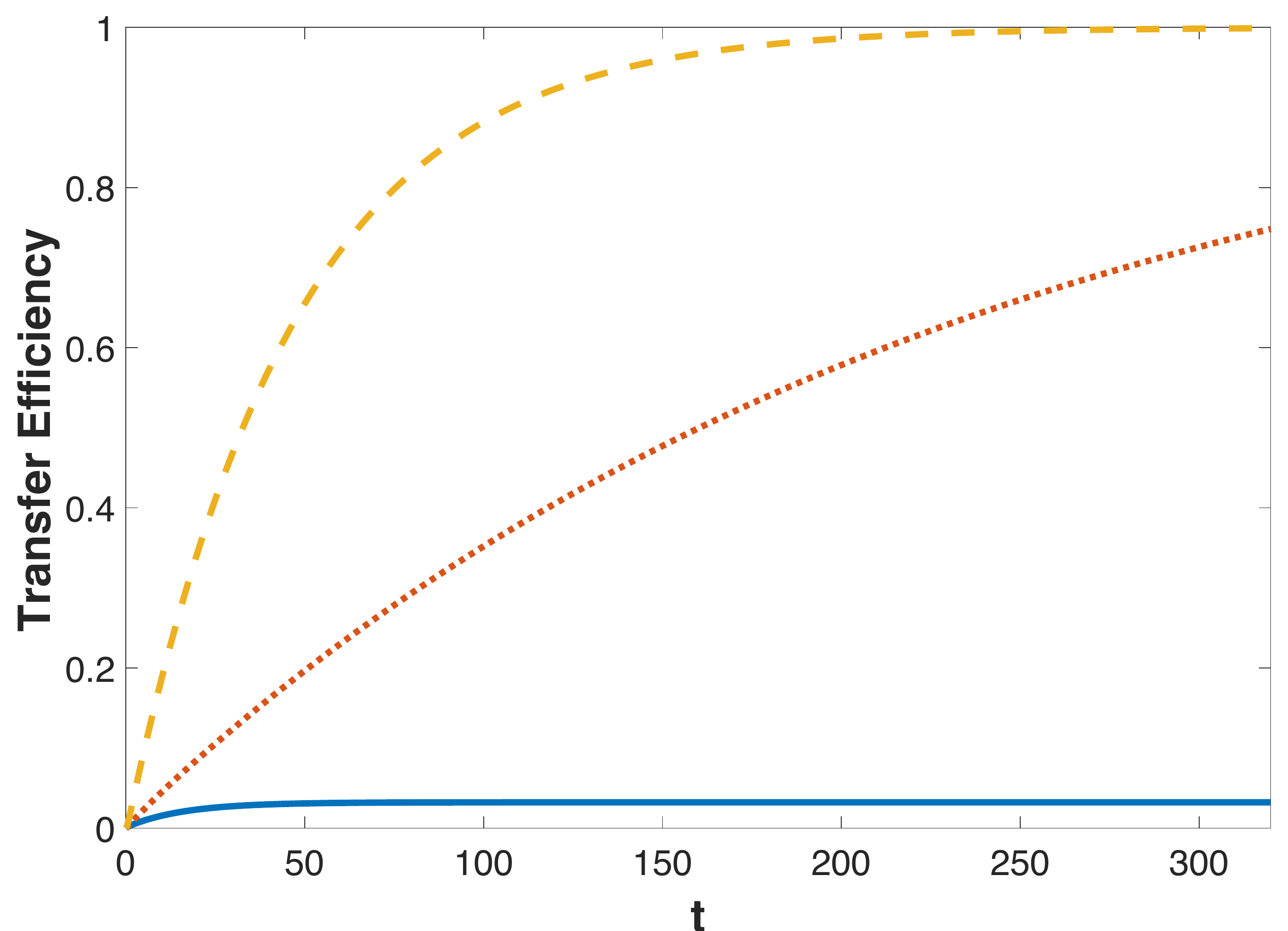}
\qquad
\includegraphics[width=0.478\textwidth]{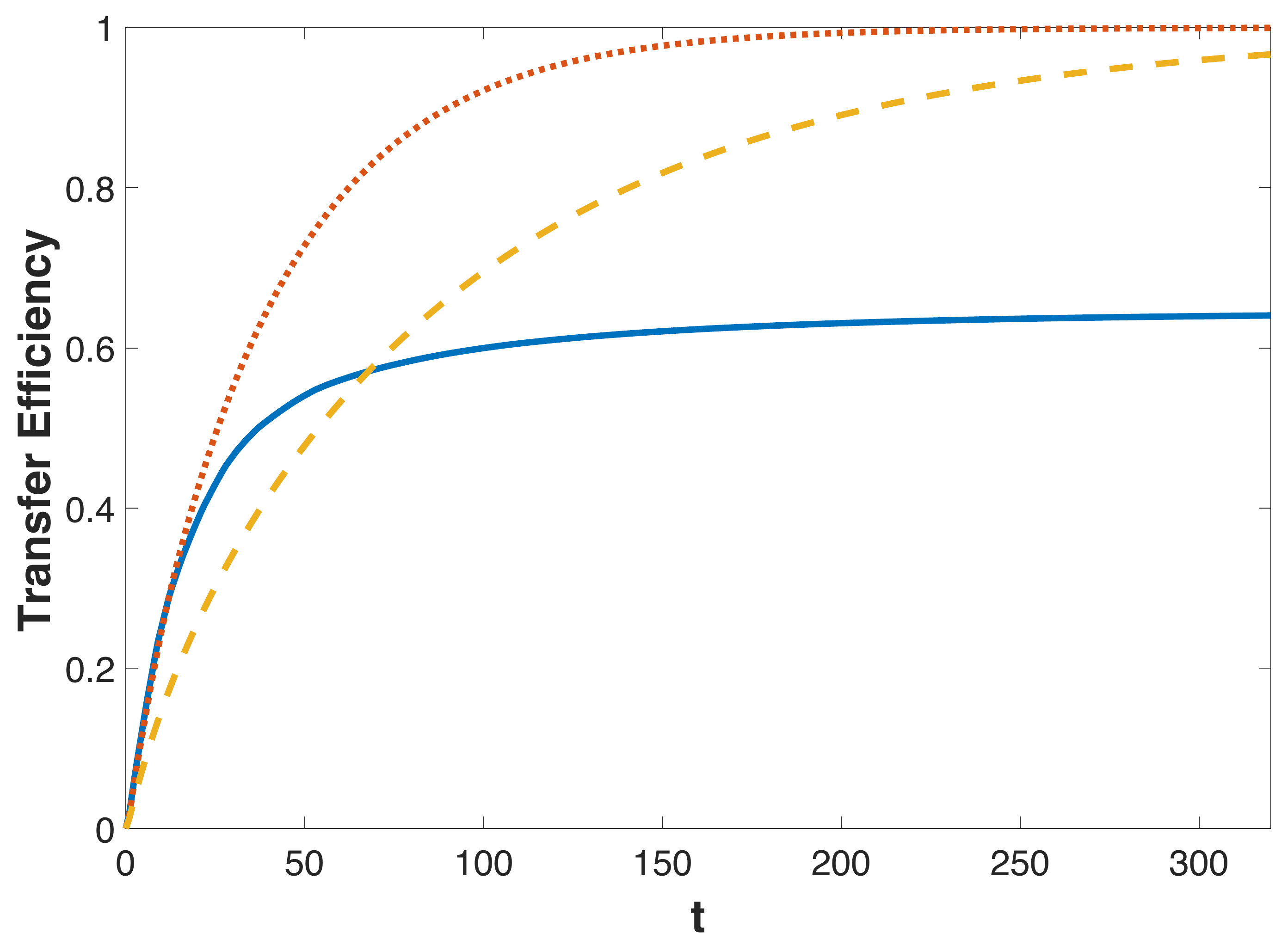}
\par\end{centering}
\caption{Transfer efficiency as a function of time, in the case of FCN (left) and a $4$x$8$-cylinder (right), both with $N=32$ nodes. Different noise conditions are shown: no dephasing (continuous line), classical dynamics (dashed) and the optimal mixing rate between these two last regimes (dot). The transfer efficiency has been averaged over all possible input and output nodes. For simplicity, we have considered $\Gamma_{j}=0$ for $j=1\dots N$ and $\Gamma_{N+1}=1$. }
\label{fig:fcn_cyl}
\end{figure*}

The application of control fields is often not desirable, however.
A static network that has high transfer efficiency is generally simpler
to implement. Since the breaking of symmetry can lead to enhanced
transfer, one can indeed add randomness or dissipative dynamics
to break symmetry and assist transport \citep{Caruso_2009}.

\subsection{Disorder And Dephasing\label{subsec:Disorder-And-Dephasing}}

For a FCN of size $N$, \citet{Caruso_2009} find that
the probability of transfer is
\begin{align}
p_{sink}\left(\infty\right) & =\dfrac{1}{N-1},
\end{align}
\emph{i.e.}, for large networks the transfer is very small. In fact, such
perfectly coherent networks are even worse than classical networks
with incoherent hopping which have \emph{complete }transfer in the
limit $t\rightarrow\infty$. The poor transfer can be seen as being
due to the large size of the dark subspace, given the network symmetries intrinsic
in the complete graph with identical nodes --- in fact, it has the largest possible dark subspace of dimension $N-2$ for a network of $N$ nodes. By introducing \emph{static disorder} to $D$ local node energies, the dark subspace reduces in
size and the probability increases to
\begin{align}
p_{sink}\left(\infty\right) & =\dfrac{1}{N-D-1}.
\end{align}
Hence, for a FCN with $D=N-2$ different (disordered)
node energies, $p_{sink}\left(\infty\right)=1$. Any initial state
has no component in any remaining invariant subspace \citep{Caruso_2009}.
Static disorder can also make transfer more robust against dissipation/noise
in the weak dissipation regime \citep{Wu_2013}.

Local dephasing on the network nodes has a very similar effect. If
there is local dephasing on \emph{all} nodes then the dark subspace
can vanish, and $p_{sink}\left(\infty\right)\rightarrow1$. In the special case of FCN the best method to obtain a unity transfer efficiency in short times is to apply strong dephasing, which leads to complete lack of coherence and so to a classical dynamics; this is due to the large size of the dark subspace, as discussed before. Instead, other networks need an interplay between quantum coherence and dephasing to destroy the invariant subspace and to obtain the same performance of FCN in the classical regime \citep{Caruso_2009, Caruso_2014}. These two different behaviours are shown in Fig. \ref{fig:fcn_cyl}, for a FCN and for a cylinder, both with $N=32$ nodes.

Dephasing
also leads to line broadening, \emph{i.e.} another way to view the enhanced
transport is due to the stronger overlap between excitation lines
of the interacting nodes \citep{Caruso_2009}. With the combination
of dephasing and static disorder, static disorder is only advantageous
when dephasing is weak. When noise (dissipation or dephasing) is too strong, quantum Zeno phenomena occur and the dynamics is frozen \citep{Caruso_2009, Caruso_2014, Wu_2013, Viciani_2016}: this may be exploited for storage.

\section{Graph theorems \label{sec:Analysis-Of-Dark}}
For uniform site energies and coupling rates, \emph{i.e.} $H=A$, we can apply two theorems from graph theory which ultimately give the existence of network dynamics for which there are \emph{no} dark subspaces.
The first result is based on the following theorem. Given a real symmetric matrix $A = [a_{ij}]$ of size $N$, one can always associate a weighted graph $G$ with $N$ nodes and with edges $\{i,j\}$ that have weights $a_{ij}$ for $i\neq j$.

\begin{thm}[\citet{Hassani_Monfared_2016}]
	For a given connected graph $G$ of $N$ vertices, and given a set of distinct values $\lambda_1, \lambda_2,\ldots,\lambda_N$, there exists a real symmetric matrix $A$ whose graph has the same topology as $G$ and whose eigenvalues are $\lambda_1, \lambda_2,\ldots,\lambda_N$, such that none of the eigenvectors of $A$ have a zero entry.   \label{thm:Monfared}
\end{thm}

By the above theorem, if we have some given underlying connected topology given by graph $G$, then we can find a set of weightings for the edges---interactions between the different nodes---such that the corresponding adjacency matrix $A(G)$ of the graph has distinct eigenvalues, and all the corresponding eigenvectors have no zero entry. With such dynamics, there is no dark subspace on the network relative to any target node.

\begin{cor}
	For any given underlying connected graph $G$, there exists Hamiltonian dynamics on the graph for which there is no dark subspace.\label{cor:No-dark-subspace}
\end{cor}

Real networks tend to have eigenvalues with higher multiplicities
(degeneracy) than comparable randomly generated networks \citep{Marrec_2017}. However, if we are able to change the interactions between the nodes that are joined, using, for example, a combination of a different underlying Hamiltonian, control fields, and disorder and noise, we can eliminate the dark subspace altogether and achieve perfect energy transfer.
In addition, our next result ultimately states that we do not even need to consider weighting the edges if the graph in question is sufficiently large.

Erd\H{o}s-R\'{e}nyi graphs $G\left(N,p\right)$ have $N$ nodes, in which
any edge between any two nodes has some probability $p$ of being
there \citep{Erdos_1959}. These graphs are very likely\footnote{A property $P$ of a graph holds \emph{asymptotically almost surely} for $G\left(N,p\right)$
if the probability of $P$ being true goes to one as $N\rightarrow\infty$.} to be disconnected if $p<\ln\left(N\right)/N$ \emph{i.e.} if the probability
of edges is sufficiently low \citep{Albert_2002}. Note that for $p\neq0, 1$, the set of all  $G\left(N,p\right)$ graphs is equivalent to the set of all graphs, since any graph will be an instance of an Erd\H{o}s-R\'{e}nyi graph. Given this fact, we can use the following theorem to subsequently make a statement about all asymptotically large graphs:

\begin{thm}[\citet{ORourke_2016}]
	 A graph $G\left(N,\frac{1}{2}\right)$
	 is controllable with probability at least $1-CN^{-\alpha}$, for any
	 $\alpha$, where $C>0$.\label{thm:A-random-graph-almost-surely}
\end{thm}

\begin{figure*}[!t]
\begin{centering}
\includegraphics[width=0.478\textwidth]{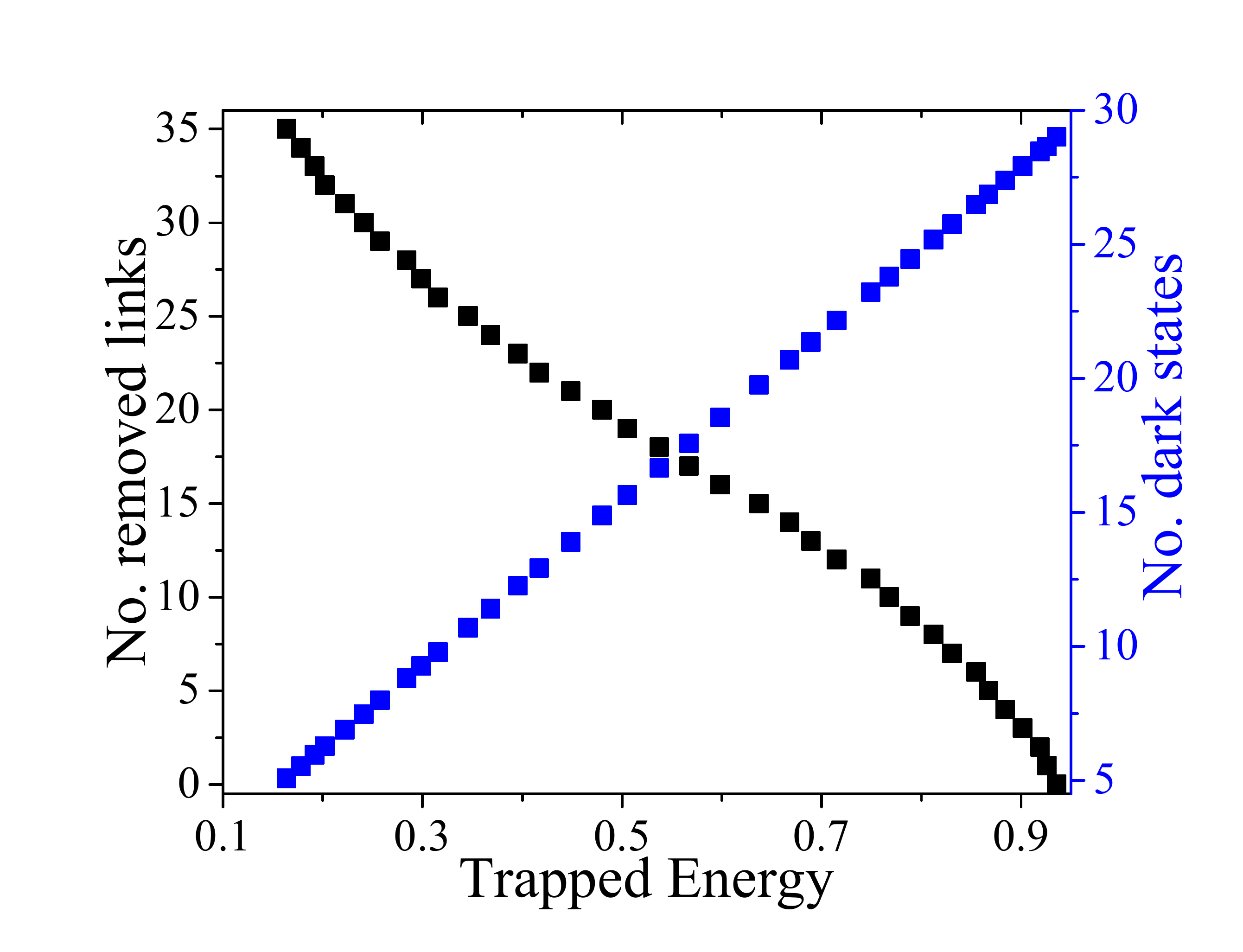}
\qquad
\includegraphics[width=0.478\textwidth]{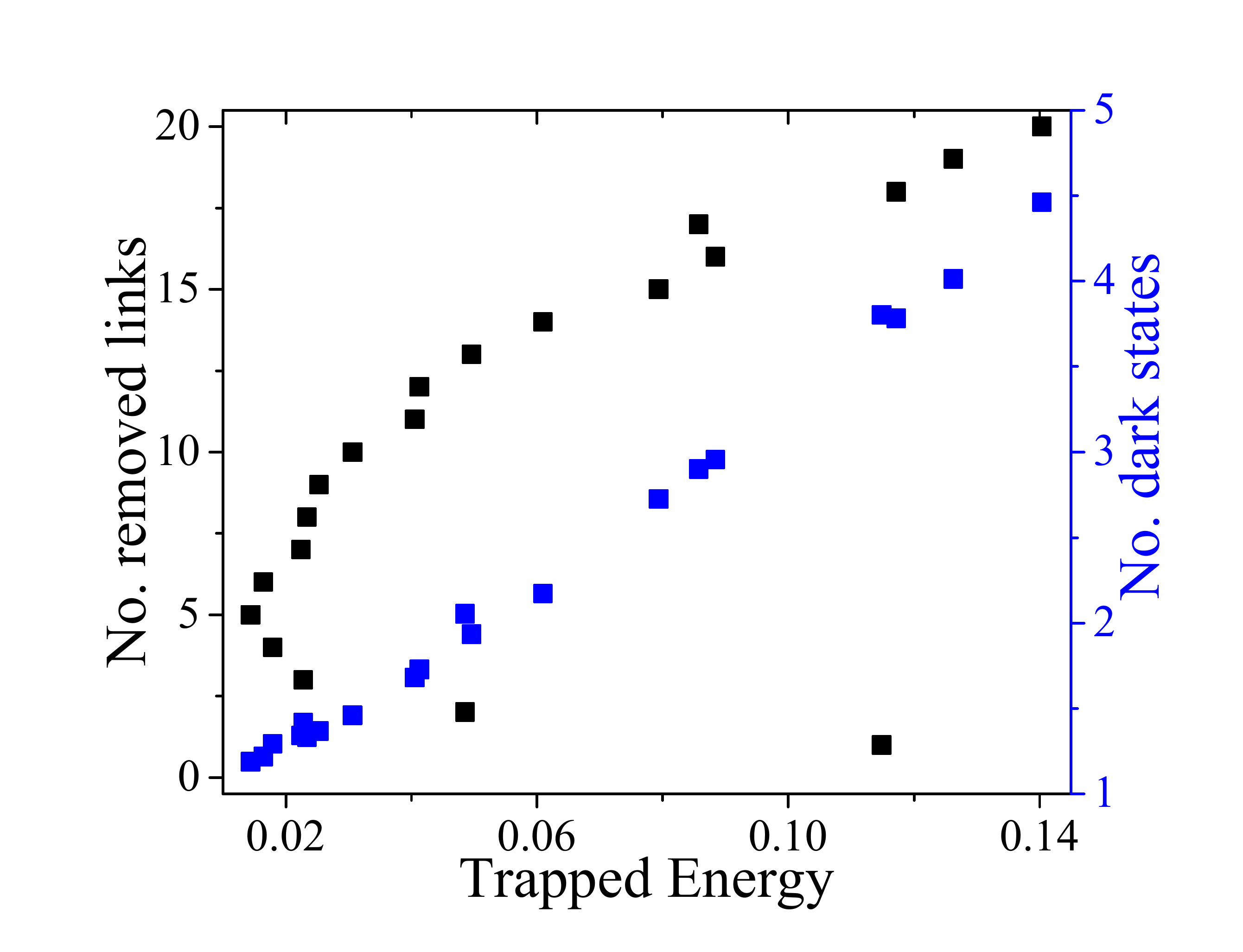}
\par\end{centering}
\caption{Relationship between the number of removed links, the energy trapped in the dark subspace and the number of dark states in case of FCN (left), and $4$x$8$-cylinder (right); both have $N=32$ nodes.}
\label{fig:fcn_cyl,trapener-NLR-DS}
\end{figure*}

 This theorem was conjectured by \citet{Godsil_2012}  (see also \cite{Godsil_2010}), and proven by  \citet{ORourke_2016}. The notation of controllability is the 	same as that introduced in Subsection \ref{subsec:Control-Fields}, \emph{i.e.}, the graph is controllable if the dynamics (determined by the adjacency matrix, which is equivalent to the Hamiltonian) can evolve any state into any other state on the graph. Stated in another way,  Theorem \ref{thm:A-random-graph-almost-surely} implies that the relative number of controllable graphs to any graph tends to one as $N\rightarrow \infty$. By picking a very large graph at random, it is almost surely controllable, and thus almost surely has no dark states.
	
\begin{cor}
A connected graph $G$ of size $N$, with Hamiltonian dynamics given by the adjacency matrix, asymptotically almost surely has no dark subspace as $N\rightarrow\infty$.
\end{cor}

Hence almost surely, energy transfer on large graphs will happen perfectly if we allow for time $t\rightarrow\infty$, without requiring the addition of further controls or different interaction strengths between the nodes.

\section{Application to light-harvesting\label{sec:Environment-Coupling}}

Real quantum networks are always subjected to noise. However, environmental interaction can \emph{enhance} transport through a dissipative network.
This is true even in classical mechanics, but via physically different mechanisms (\emph{e.g.}, stochastic
resonance \citep{Gammaitoni_1998}). Besides, quantum mechanically, noise can
maintain and even generate quantum coherence and entanglement \citep{Plenio_2002,Huelga_2007,Rivas_2009,Braun_2002,Benatti_2003}.

The transport of excitations in light-harvesting complexes has attracted
much interest in the last decade. Light-harvesting complexes, or antenna systems, are networks
composed of chromophores absorbing photons and transporting the created electronic excitations to the reaction centre (the target node). In particular, in the simplest light-harvesting complex, known as Fenna-Mathews-Olson (FMO) complex, found in green sulphur bacteria, experimental evidence strongly suggests that quantum coherence features play a crucial role during the energy
transport process \citep{Engel_2007,Panitchayangkoon_2010,Panitchayangkoon_2011}.
%\begin{figure}
%	\begin{centering}
%		\includegraphics[width=0.56\textwidth]{Caruso_2009.jpeg}
%		\par\end{centering}
%	\caption{Modelling of $p_{sink}$ for the FMO complex by \citet{Caruso_2009}
%		as a fully connected network with unequal edge weights and site energies.
%		The black line corresponds to the noiseless scenario, and the red
%		line is at optimal local dephasing showing enhanced transport. Figure
%		from Ref. \citep{Caruso_2009}. \label{fig:Modelling-of-FMO-Caruso}}
%\end{figure}
Theoretical studies show that the additional presence of dephasing noise is needed to describe the observed transport efficiency of almost $100\%$ \citep{Caruso_2009,Mohseni_2008, Plenio_2008, Olaya-Castro_2008, Caruso_2010_LHC, Chin_2010}.
%From the shape of the curves in Fig. \ref{fig:Modelling-of-FMO-Caruso},
%there is an initial rapid increase in $p_{sink}$, followed by a slower
%increase due to the state on the FMO network being in an \emph{approximate}
%dark subspace, which eventually will transfer to the sink.

A more recent example of experimental evidence where it is possible to obtain an optimal transport combining quantum coherence and noise is described in Ref. \citep{Park_2016}. Particularly, a light-harvesting antenna system has been realized with a biological material, the M13 virus, and a chromophore network has been created on its filaments. Two versions of this system have been genetically planned: one with a network made of weakly coupled chromophores, and the other one with reduced inter-chromophoric distance, causing clusters of strongly coupled chromophores. In this second version, involving coherent and incoherent features, they have observed a remarkable improvement of both transport speed and diffusion length of the electronic excitation. The average chromophoric distance was exploited to study and control the optimal mixing rate between coherence and noise.
Here, the environment assists the transport by suppressing the dark subspaces or inducing interaction between them and other states, causing ultimate leakage into the sink \citep{Caruso_2009, Wu_2013}.

In this paper, our particular choice of the cylinder graph for quantum transport simulations is indeed inspired by the topology of this virus structure.

\section{Topology robustness\label{sec:robustness}}

\begin{figure*}[!t]
\begin{centering}
\includegraphics[width=0.478\textwidth]{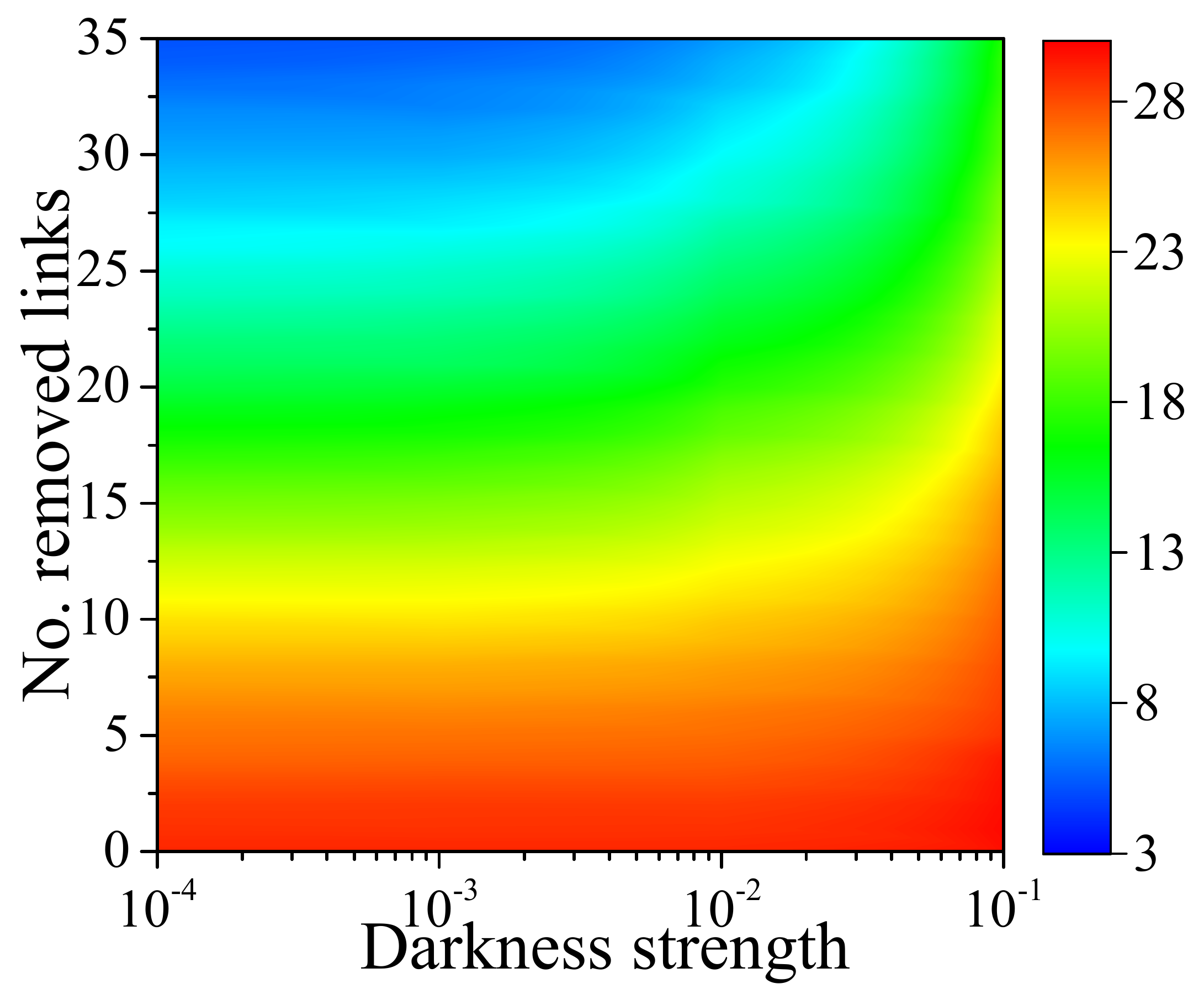}
\qquad
\includegraphics[width=0.478\textwidth]{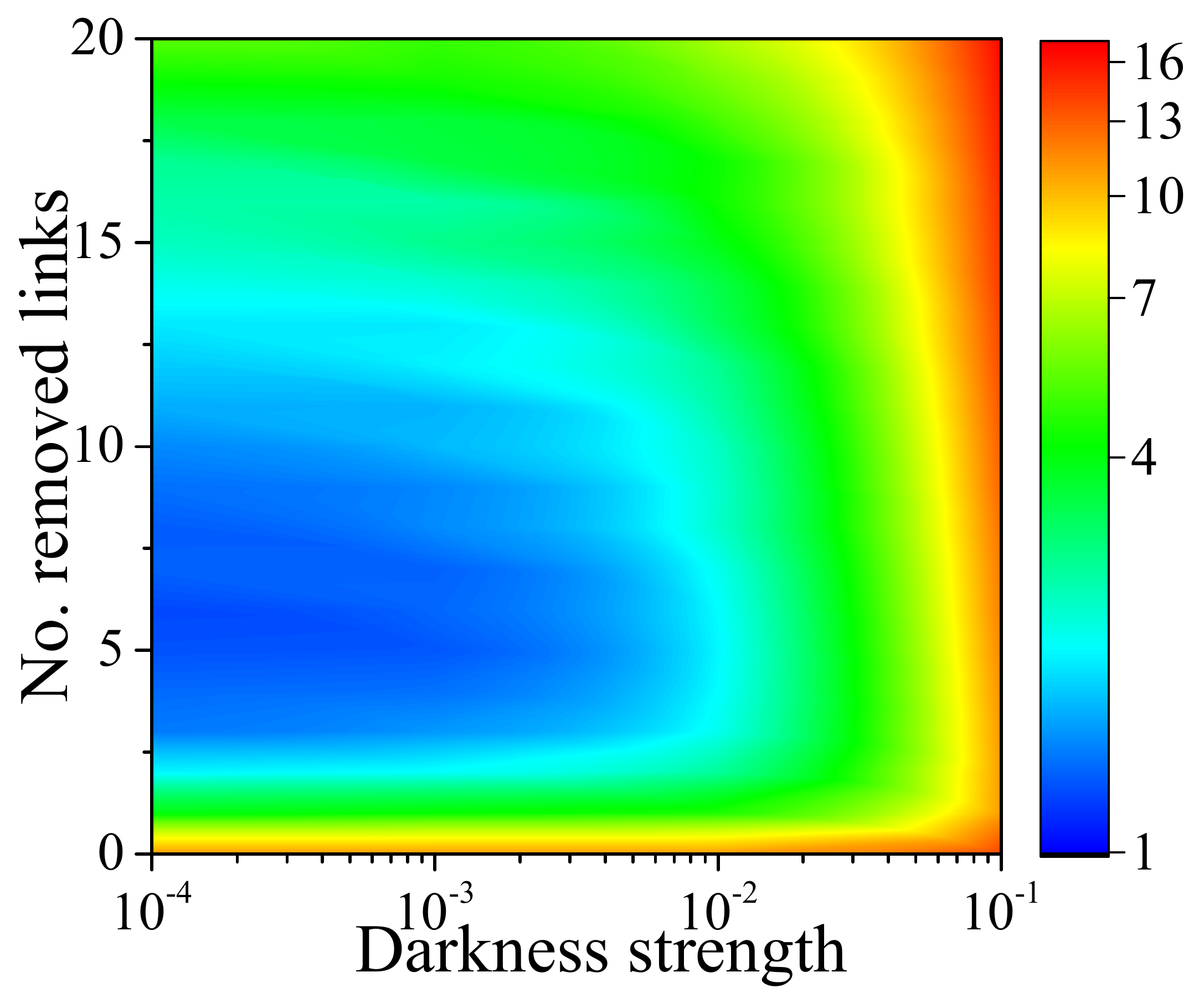}
\par\end{centering}
\caption{Number of dark states as a function of the number of removed links and the darkness strength, for FCN (left) and a  $4$x$8$-cylinder graph (right), with $32$ nodes. A similar qualitative behaviour is observed for larger networks.}
\label{fig:fcn_cyl,darknstren-NRL-NDS}
\end{figure*}

Inspired by Theorem \ref{thm:Monfared}, we have implemented some numerical simulations that randomly remove a specific number of links in the network. This approach enables us to study both the effect on dark subspaces and to mimic a real condition that could happen in presence of perturbations (\emph{e.g.}, material defects).
Removing links is beneficial to the FCN because it reduces the dark subspace dimension, hence also reducing the amount of trapped energy. In contrast, the cylinder graph benefits from link deletion only up to a small percentage of removed links (about 5\% of the total); when this percentage grows another dark subspace appears again and the transport gets worse. As we can see in Fig. \ref{fig:fcn_cyl,trapener-NLR-DS}, the energy trapped grows linearly with the number of dark states for both the FCN and cylinder networks. As the number of removed links grows, the energy trapped on the FCN network monotonically decreases, whilst the energy trapped on the cylinder network decreases initially and then increases again. This last behaviour is probably due to the appearance of new symmetries, hence new dark states appear.

However, although the deletion of links is a good method to reduce the number of dark states, it is not sufficient in reducing the presence of quasi-dark states, since the latter are more persistent. In Fig. \ref{fig:fcn_cyl,darknstren-NRL-NDS} the plotted quantity is the number of dark states and quasi-dark states as a function of the darkness strength and of the number of deleted links. Note that it turns out to be more difficult to destroy quasi-dark states by means of removing links. Moreover, in agreement with Fig.\ref{fig:fcn_cyl,trapener-NLR-DS}, after removing too many links the appearance of new dark states can occur, as shown in the right panel of Fig. \ref{fig:fcn_cyl,darknstren-NRL-NDS} -- see its left panel for FCN as comparison. 

\begin{figure}[b!]
\begin{centering}
\includegraphics[width=\columnwidth]{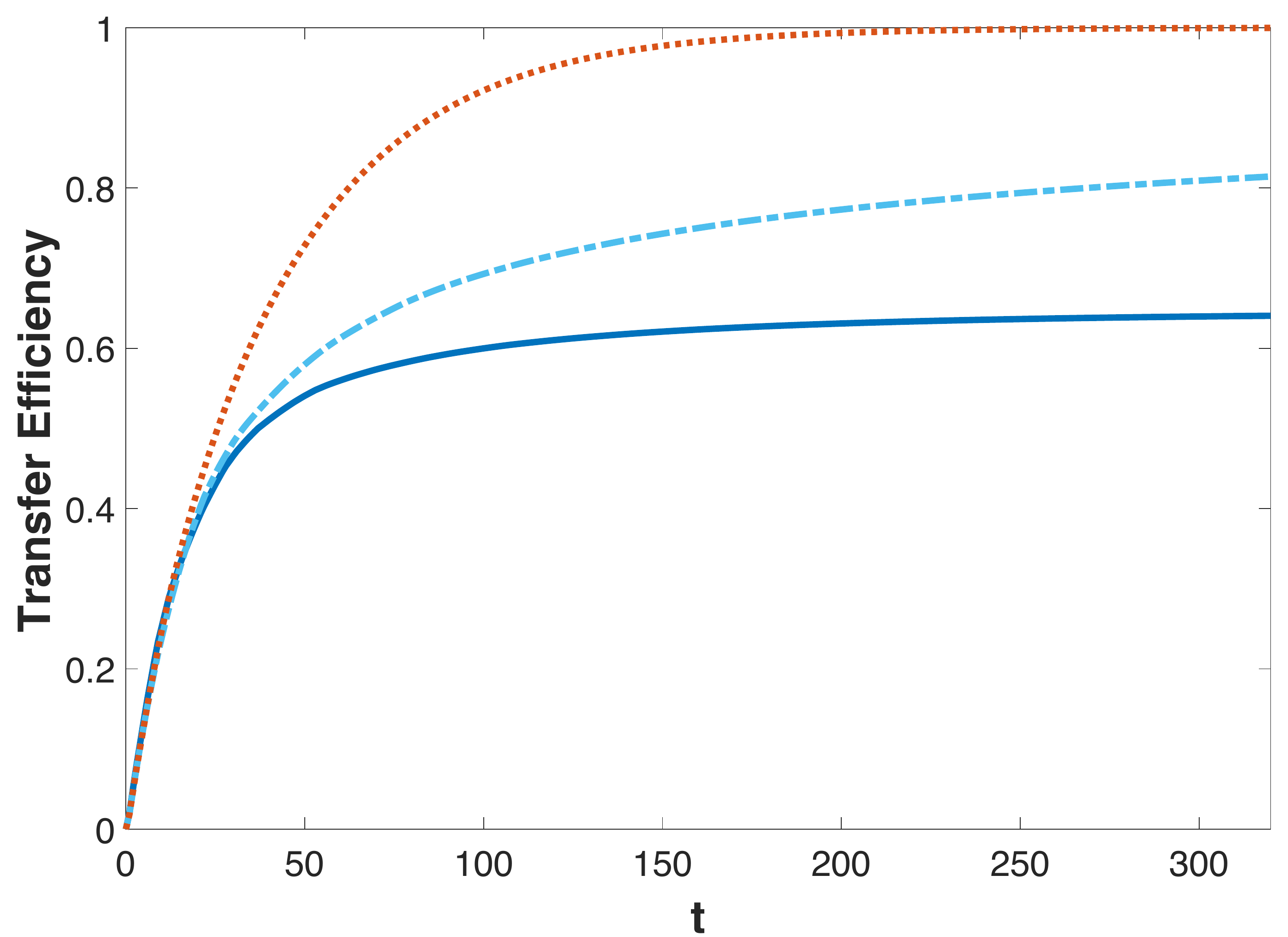}
\par\end{centering}
\caption{Time evolution of the transfer efficiency for a  $4$x$8$-cylinder of $N=32$ nodes in three different conditions: optimal dephasing noise with no removed links (dot line), coherent dynamics  with no removed links (continuous), and coherent dynamics but with $5$ removed links (dot-dashed). Each transfer efficiency has been averaged over all possible input and output states, and with $\Gamma_{j}=1$ only for $j=N+1$, as in Fig. \ref{fig:fcn_cyl}.}
\label{fig:cyl_cylNLR3p0,01}
\end{figure}

In this context, dephasing noise opens up additional pathways from the initial node to the final one and therefore suppresses both dark states and quasi-dark states. The presence of noise is more effective than links deletion for transport improvement. Indeed, in Fig. \ref{fig:cyl_cylNLR3p0,01} we have plotted the time evolution of transfer efficiency of a cylindrical topology, comparing the case of no removed links with the one of an optimal number of removed links (corresponding to the minimum of the energy trapped --- see Fig. \ref{fig:fcn_cyl,trapener-NLR-DS}). As already discussed above, without links deletion we have a dark subspace obstructing electronic excitation from reaching the sink. Then, removing $5$ links allows us to obtain $p_{sink}\left(\infty\right)=1$. If the aim is instead the achievement of an optimal fast transport, dephasing noise plays a crucial role: in fact $p_{sink}$ reaches unity in a much shorter time scale (dot line in Fig. \ref{fig:cyl_cylNLR3p0,01}).

\begin{figure}[!t]
\begin{centering}
\includegraphics[width=0.5\textwidth]{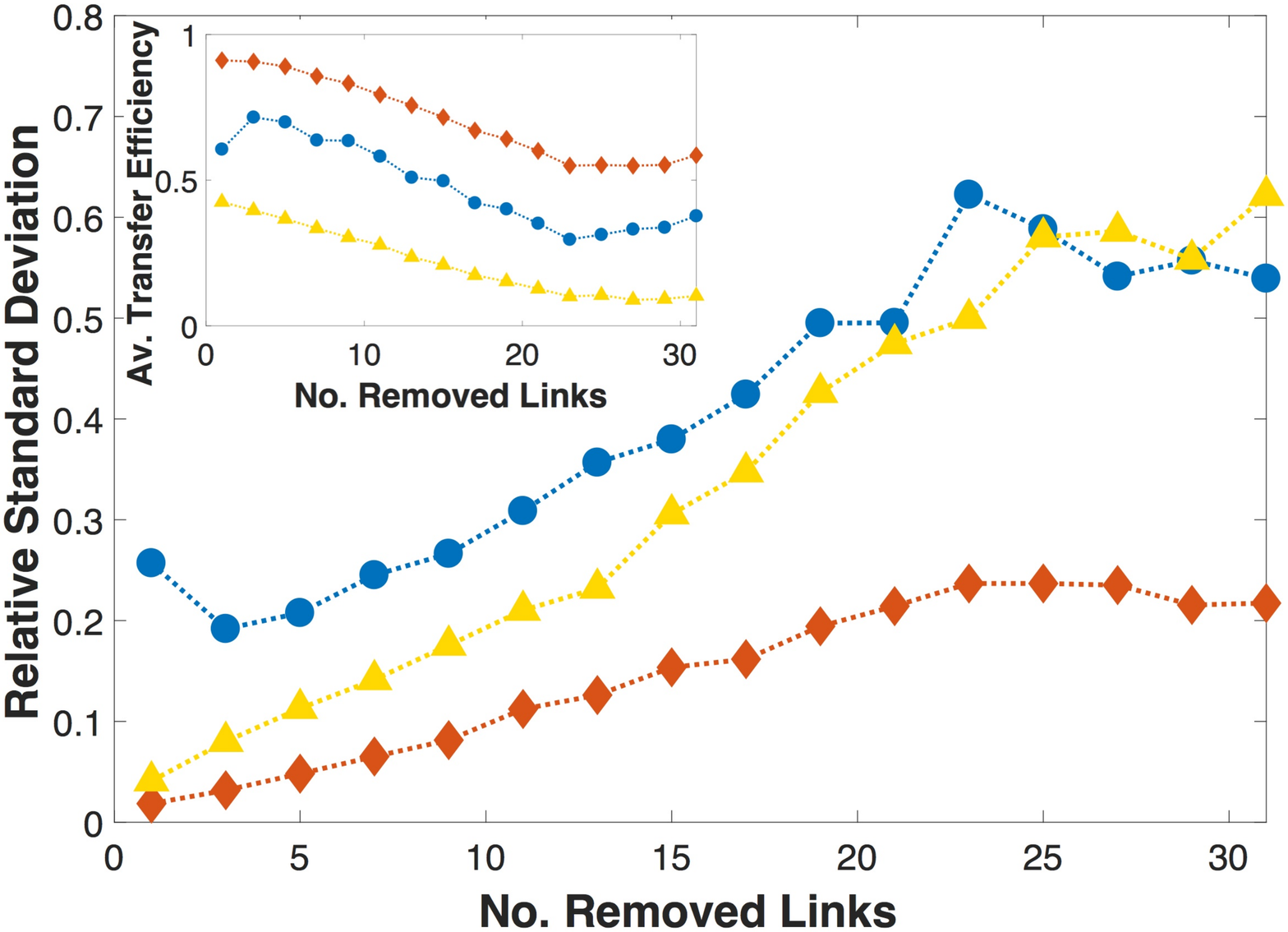}
\par\end{centering}
\caption{Relative standard deviation of transfer efficiency as a function of the number of removed links for a  $4$x$8$-cylinder with $N=32$ nodes, with fixed input and output nodes (at opposite ends) and $\Gamma_{N+1}=1$, and corresponding to a sample of $200$ different geometries. The corresponding averaged transfer efficiency is shown in the inset.}
\label{fig:cyl_rsd-NRL}
\end{figure}

 Indeed, noise-assisted transport is characterized not only by a reduced time scale for the transmission, but also by the robustness against possible changes of the underlying topology, as discussed in \cite{Caruso_2014}. By varying the geometry and adding the right amount of noise, a very good transport performance is guaranteed. This does not occur in the fully coherent and incoherent cases, where the transfer efficiency quickly decreases, as it can be seen in the inset of Fig. \ref{fig:cyl_rsd-NRL}. This remarkable robustness is present in the regime of noise-assisted transport, as shown by the smaller dispersion around the optimal efficiency with respect to the fully coherent and incoherent regimes. Finally, let us point out that the minimum of the relative standard deviation and the maximum of the average of the transfer efficiency in the coherent case (corresponding to $5\%$ of removed links) is a further sign of dark subspace suppression.

\section{Conclusions\label{sec:Conclusion}}

The dark side of quantum networks is an antagonist to optimal energy
transfer. Different tools can be employed to deal with the dark subspaces:
we can avoid them using smart initialisation, or suppress and destroy
them by breaking the network symmetries through the use of control
fields, noise, or disorder. Indeed, dark subspaces have a deep
connection with topological symmetries, and can grow in size
on more symmetric networks (associated to more degenerate adjacency matrices). The FCN network, for example, has the most symmetries possible on a network
and hence the largest dark subspace. At the same time, the FCN network
also responded most favourably to dark space suppression tools as
opposed to the less symmetric cylinder graph. Whilst the dark subspace
has been defined in relation to the eigenstates of the Hamiltonian
describing the dynamics on the network, the framework of the dark
subspaces could also be generalised to include other features, such
as impurities that trap and cause decay of energy on
the network \citep{Agliari_2010}, and to Lindbladian eigenstates in
more generality.
The best method to get optimal transport would depend on the function of the device we want to plan: if the goal is the unity of $p_{sink}$ without time limits, then designing a proper weighted network could be the solution (assuming that it is within our engineering ability); if short times and performance robustness are crucial (as it is usually the case), then the introduction of noise in the dynamics is required. Given that noise is unavoidable in most realistic systems, this implies
that we generally do \emph{not} need to eradicate all noise to achieve
optimal transport\textemdash we just need to be able to control it
to some degree.

Besides, we found that a network does not have any \emph{truly} dark states,
if the interactions can be tuned to achieve full controllability although this may not be quite feasible experimentally.
If the interactions
can be engineered, then this is advantageous in two ways: first, no excitation is truly trapped on the network, hence we can always be
sure that full transfer will eventually occur; second, there will be  ``temporary'' dark states that could be used as energy storage. Furthermore, sufficiently large graphs  almost surely have no dark states, implying that as our quantum networks grow in size (\emph{i.e.}, as the particular quantum technology grows in size), we are very likely
to not require extensive interaction engineering to ensure full transport.

These results allow one to move further in understanding and enhancing
state transfer on quantum networks \citep{Christandl_2004,Bose_2009,Kendon_2011}. These results
can also be employed to understand
other quantum processes such as electron transfer, and to designing
solar energy devices (\emph{e.g.}, inspired
by the energy transfer networks in photosynthetic complexes), and potential quantum thermal devices.

\section{Acknowledgements}

We would like to thank Joshua
Lockhart, Yasser Omar, Danial Dervovic, Bryan Shader, Gabriel Coutinho and Stefano Gherardini for useful discussions. This work was supported by the EPSRC Centre for Doctoral Training in Delivering Quantum Technologies [EP/L015242/1]. F. C. was also financially supported from the Fondazione CR Firenze, through the project Q-BIOSCAN; S. S. was financially supported from the Royal Society, EPSRC, Innovate UK, BHF and NSCF.

\bibliographystyle{apsrev4-1}
\bibliography{quantum_transport}

\end{document}